\documentclass[aps,pra,groupedaddress,twocolumn,superscriptaddress]{revtex4-1}

\usepackage[utf8x]{inputenc}

\usepackage{color}

\usepackage{bbm} 

\usepackage{amsfonts,amsmath,amssymb,stmaryrd}

\usepackage{graphicx}

\usepackage{subfigure}  % use for side-by-side figures

\usepackage{bbm} 

\usepackage{hyperref}

\usepackage{epsfig}

\usepackage{verbatim}

\renewcommand{\l}{\left(}

\renewcommand{\r}{\right)}

\newcommand{\Ys}{\prod_{j<k}(z_j-z_k)}

\newcommand{\CF}[1]{\hspace{-0.12cm}~^{#1}\text{CF}}

\newcommand{\CB}[1]{\hspace{-0.12cm}~^{#1}\text{CB}}

\newcommand{\CP}[1]{\hspace{-0.12cm}~^{#1}\text{CP}}

\newcommand{\LLL}{\text{LLL}}

\newcommand{\LN}{\text{LN}}

\newcommand{\ket}[1]{|#1\rangle}

\newcommand{\cl}{\text{cl}}

\begin{document}
\title{Fractional quantum Hall physics with ultracold Rydberg gases in artificial gauge fields}

\author{F. Grusdt}
\affiliation{Department of Physics and research center OPTIMAS, University of Kaiserslautern, Germany}
\affiliation{Graduate School Materials Science in Mainz, 67663 Kaiserslautern, Germany}
\author{M. Fleischhauer}
\affiliation{Department of Physics and research center OPTIMAS, University of Kaiserslautern, Germany}


\date{\today}

\begin{abstract}
We study ultracold Rydberg-dressed Bose gases subject to artificial gauge fields in the fractional quantum Hall (FQH) regime. The characteristics of the Rydberg interaction 
gives rise to interesting many-body ground states different from standard FQH physics in
the lowest Landau level (LLL). The non-local but rapidly decreasing interaction potential favors crystalline ground states 
for very dilute systems. While a simple Wigner crystal becomes energetically favorable compared to the Laughlin liquid for filling fractions $\nu<1/12$, a correlated crystal of 
composite particles emerges already for $\nu \leq 1/6$ with a large energy gap to the simple
Wigner crystal.
The presence of a new length scale, the Rydberg blockade radius $a_B$, gives rise to a bubble crystal phase for $\nu\lesssim 1/4$ when the average particle distance 
becomes less than $a_B$, which describes the region of saturated, almost constant interaction potential. For larger fillings indications for strongly correlated cluster liquids are found. 
\end{abstract}

\maketitle
%%%%%%%%%%%%%%%%%%%%%%%%%%%%%%%%%%%%%%%%%%%%%%%%%%%%%%%%%%%%%%%%%%%%%%%%%%%%%%%%%%%%%%
%%%%%%%%%%%%%%%%%%%%%%%%%%%%%%%%%%%%%%%%%%%%%%%%%%%%%%%%%%%%%%%%%%%%%%%%%%%%%%%%%%%%%%

%%%%%%%%%%%%%%%%%%%%%%%%%%%%%%%%%%%%%%%%%%%%%%%%%%%%%%%%%%%%%%%%%%%%%%%%%%%%%%%%%%%%%%
\section{Introdcution}
%%%%%%%%%%%%%%%%%%%%%%%%%%%%%%%%%%%%%%%%%%%%%%%%%%%%%%%%%%%%%%%%%%%%%%%%%%%%%%%%%%%%%%

In the context of the fractional quantum Hall effect (FQHE) interesting many body ground states exists, some of which carry fractional and non-abelian braiding statistics \cite{HALPERIN1984,AROVAS1984,MOORE1991,Nayak1996,Bonderson2011}. This makes them 
ideal candidates for a topological quantum computer \cite{Kitaev2003,Nayak2008} and gives access to topological quantum phases 
\cite{THOULESS1982}. For an experimental study 
a high degree of control is required and quasi-2D ultracold gases of atoms in artificial magnetic fields have 
been suggested as potential candidates \cite{Cooper2001}. Effective magnetic fields are generated either by rotation \cite{Cooper2008,Fetter2009},
employing light-induced gauge potentials \cite{Juzeliunas2004,Juzeliunas2006,Dalibard2011,Lin2009,Cooper2011},
or in lattices with complex hopping amplitudes \cite{Jaksch2003,Aidelsburger2011}.
When a weakly interacting superfluid Bose gas is rotated  sufficiently fast an ordered vortex-lattice forms
\cite{Madison2000,Abo-Shaeer2001} and the lowest Landau level (LLL) regime can be reached \cite{Schweikhard2004}. Although for bosonic atoms there is no integer quantum Hall effect
based on the Pauli principle, repulsive interactions can give rise to highly correlated FQH states for fillings $\nu \lesssim 6$ \cite{Cooper2001} when only the LLL is occupied \cite{Wilkin1998,Cooper1999,Regnault2003}. However, despite the experimental progress
strongly correlated phases have not been realized yet.
In part this can be attributed to the rather small interaction energies in atomic gases. In the present paper we propose an alternative approach using $1/r^6$ van-der-Waals (vdW) interactions in Rydberg states. 
The associated energies can be orders of magnitude larger
than those achievable with contact or magnetic dipole-dipole interactions 
\cite{Saffman2010} making Rydberg interactions an ideal candidate for the realization
of FQH physics. It is also possible to hybridize Rydberg atoms with light in 
form of so-called Rydberg polaritons \cite{Peyronel2012}, which opens the possibility to realize
FQH physics with photons. Here artificial gauge fields can be generated either
by rotation of a suitable medium \cite{Otterbach2010} or using wave-guide lattice structures \cite{Hafezi2011}.

In the present paper we theoretically analyze the phase diagram of particles with a
Rydberg interaction in the LLL using exact diagonalization (ED) as well as variational methods.
Similar to dipolar gases with $1/r^3$ interaction, which have been extensively studied in the past 
\cite{Osterloh2007,Baranov2005,Rezayi2005}, we find a transition from Laughlin (LN) liquids \cite{LAUGHLIN1983} 
to crystalline ground states for very dilute systems, however with
an extended filling region where only composite particles can crystallize at zero temperature.
In contrast to dipolar gases the difference in variational energy between composite crystals and 
simple Wigner crystals is rather large. 
Additionally the presence of a new length scale, the so-called Rydberg blockade radius $a_B$ gives rise to a 
clustering mechanism supporting bubble crystal ground states for fractional fillings 
$\nu\le 1/4$. For larger fillings we find in the regime of large blockade radii indications for very 
interesting cluster liquids.

%%%%%%%%%%%%%%%%%%%%%%%%%%%%%%%%%%%%%%%%%%%%%%%%%%%%%%%%%%%%%%%%%%%%%%%%%%%%%%%%%%%%%
%%%%%%%%%%%%%%%%%%%%%%%%%%%%%%%%%%%%%%%%%%%%%%%%%%%%%%%%%%%%%%%%%%%%%%%%%%%%%%%%%%%%%
\section{Rydberg dressed atoms and pseudopotentials in the LLL}
%%%%%%%%%%%%%%%%%%%%%%%%%%%%%%%%%%%%%%%%%%%%%%%%%%%%%%%%%%%%%%%%%%%%%%%%%%%%%%%%%%%%%
%%%%%%%%%%%%%%%%%%%%%%%%%%%%%%%%%%%%%%%%%%%%%%%%%%%%%%%%%%%%%%%%%%%%%%%%%%%%%%%%%%%%%

%%%%%%%%%%%%%%%%%%%%%%%%%%%%%%%%%%%%%%%%%%%%%%%%%%%%%%%%%

\begin{figure}[t]
\begin{center}
\epsfig{file=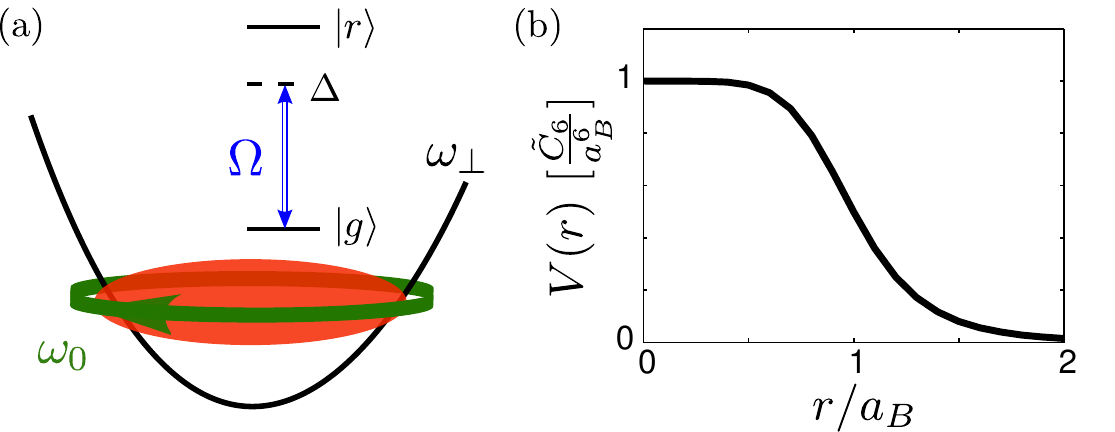, width=0.49\textwidth}
\end{center}
\caption{(color online). (a) An ultra-cold Bose gas is dressed with Rydberg-excitations $\ket{r}$ by illuminating with a far off-resonant (detuning $\Delta$) laser beam $\Omega$. 
Rotating the gas with frequency $\omega_0$ and applying a weak harmonic trap $\omega_{\perp}$ in radial and a tight one 
in the transverse direction ($\omega_z \gg \omega_{\perp}$) the quasi-2D FQH regime can be reached. (b) Interaction potential for Rydberg-dressed atoms.}
\label{fig1}
\end{figure}

Recently there has been a lot of interest in atoms excited to high Rydberg states and there has 
been considerable experimental progress to make these systems accessible
\cite{Tong2004,Singer2004,Vogt2006,Heidemann2007,Heidemann2008,Viteau2011,Schwarzkopf2011}. 
Of particular interest in the present context  are atoms excited by far-off resonant laser radiation. 
In this case, called Rydberg dressing (see fig.\ref{fig1} (a)), the atoms essentially remain in their 
ground state but show an effective interaction \cite{Henkel2010,Johnson2010} (see fig.\ref{fig1} (b))
\begin{equation}
 V(r) = \frac{\tilde{C}_6}{a_B^6 + r^6}, \qquad  a_B=\left( \frac{C_6}{2 \hbar \Delta} \right)^{1/6}.
\label{eq:effPot}
\end{equation}
For large particle separations $r$ the interaction potential is of vdW type $\sim r^{-6}$, where 
$\tilde{C}_6= \left( \Omega/ 2 \Delta \right)^4 C_6$ describes the effective interaction strength 
with $C_6$ being the bare vdW coefficient, $\Omega \ll |\Delta|$ the Rabi-frequency of the laser excitation and $\Delta$ its detuning. 
Since bare Rydberg interaction energies are typically many orders of magnitude stronger than e.g. 
magnetic dipole-dipole interactions, the excitation gap of FQH states can easily 
reach the energy separation between Landau levels (LLs).
Most importantly the interaction potential (\ref{eq:effPot}) flattens off below the blockade radius $a_B$, 
thereby defining a new characteristic length scale. We note that the
same behaviour of the interaction potential is found for Rydberg polaritons \cite{Otterbach2013},
which allows to apply many of the results obtained here to FQH physics of photons 
hybridized with Rydberg atoms.

%%%%%%%%%%%%%%%%%%%%%%%%%%%%%%%%%%%%%%%%%%%%%%%%%%%%%%%%%%%%%%%%%%%%%%%%%%%%%%%%%%%%%%%%%%%%%%%
%%%%%%%%%%%%%%%%%%%%%%%%%%%%%%%%%%%%%%%%%%%%%%%%%%%%%%%%%%%%%%%%%%%%%%%%%%%%%%%%%%%%%%%%%%%%%%%%%%%

%%%%%%%%%%%%%%%%%%%%%%%%%%%%%%%%%%%%%%%%%%%%%%%%%%%%%%%%%%%%%%%%%%%%%%%%%%%%%%%%%%%%%%%%%%%%%%%
\begin{figure}[b]
\begin{center}
\epsfig{file=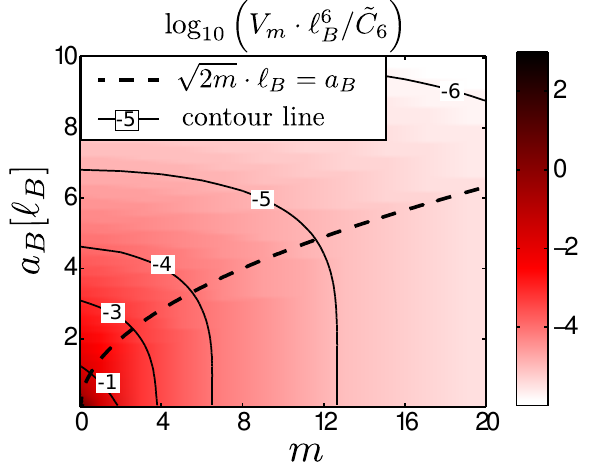, width=0.3\textwidth}
\end{center}
\caption{(color online). Contour plot of pseudopotentials $V_m$ for the potential \eqref{eq:effPot}, where $m$ was treated as continuous parameter for better illustration. Black solid lines denote contours of constant (continuous) pseudopotential.}
\label{fig1_5}
\end{figure}
%%%%%%%%%%%%%%%%%%%%%%%%%%%%%%%%%%%%%%%%%%%%%%%%%%%%%%%%%%%%%%%%%%%%%%%%%%%%%%%%%%%%%%%%%%%%%%%%%%%

To be specific we consider in the following a continuous gas of Rydberg atoms 
for which the effective magnetic field is created by rotation, noting however that
the results carry over to lattice systems with not too large flux per plaquette \cite{Hafezi2007}.
When a Rydberg-dressed Bose gas is set into rotation with angular frequency $\omega_0$ in a radial trap  of frequency $\omega_{\perp}$, the characteristic oscillator length scale is given by $\ell_c = (2 m_a \omega_{\perp}/\hbar)^{-1/2}$ \cite{Jain2007} with $m_a$ the atomic mass. The competition of the two length scales $\ell_c$ and $a_B$ leads to interesting new physics. Already in the mean-field regime where the  rotation frequency is well below the deconfinement limit ($\omega_{\perp} > \omega_0$) there is e.g. a transition from a vortex-lattice to a supersolid phase \cite{Henkel2012}. Here we are interested in the regime of strong correlations. To this end
we consider a Rydberg-dressed Bose gas in the LLL regime for fillings $\nu < 1$, assume a quasi-2D gas, neglect finite-thickness effects and use $\omega_{\perp}=\omega_0$. In the latter case $\ell_c$ is replaced by the \emph{magnetic} length $\ell_c \rightarrow \ell_B$. 
For simplicity we consider only even values of $1/\nu$ corresponding to bosonic LN states.

The interaction projected to the LLL is described by Haldane's two-particle pseudopotentials $V_m$ \cite{Haldane1983}, where the integer $m$ denotes the relative angular momentum. 
In disc geometry the pseudopotenials are given by
\begin{equation}
 V_m=\frac{1}{2^{2m+1}m!}\int_0^{\infty}dr ~ r^{1+2m}V(r) e^{-r^2/4},
\label{eq:defVm}
\end{equation}
where for bosons (fermions) only $m=0,2,4,...$ ($m=1,3,5,...$) are relevant. 

The dominant energy scale is determined
by the lowest pseudopotential $V_0$, while the higher-order terms typically fall off very fast.
For a strict point interaction in the LLL all $V_m$'s vanish identically except for $m=0$.
Tables \ref{tab:V0} and \ref{tab:V0a} list some explicit values of $V_0$ for Rb.
There we assume a fixed magnetic length $\ell_B\approx 1 \mu \text{m}$ (corresponding to $\omega_c \approx 2 \pi \cdot 130 \text{Hz}$ which is a realistic value \cite{Bretin2004}) as well as a fixed ratio $\Omega/ \Delta = 0.1$. For the $60S_{1/2}$ Rydberg state in Rb, $C_6(n=60)/2 \pi = 0.14 \text{THz} \mu \text{m}^6$ and we used the scaling law $C_6(n) \propto n^{11}$ \cite{Singer2005}. In table \ref{tab:V0} $\Delta = 2 \pi \cdot 1 \text{GHz}$ is constant and the principle quantum number of the Rydberg state $n$ is varied. In table \ref{tab:V0a} on the other hand, $n=46$ is fixed and the detuning $\Delta$ is varied. Note that by varying $\Delta$ it is possible to tune the interaction potential adiabatically.

%%%%%%%%%%%%%%%%%%%%%%%%%%%%%%%%%%%%%%%%%%%%%%%%%%%%%%%%%%%%%%%%%%%%%%%%%%%%%%%%%%%%%%%
\begin{table}[t]
\centering
\begin{tabular}{c||p{0.8cm}|p{0.8cm}|p{0.8cm}|p{0.8cm}|p{0.8cm}|p{0.8cm}|p{0.8cm}}
$n$ & $40$ & $50$ & $60$ & $70$ & $80$ & $90$ & $100$\\ \hline
$a_B ~ [\mu \text{m}]$ & $0.97$ & $1.5$ & $2.0$ & $2.7$ & $3.4$ & $4.3$ & $5.2$\\ \hline
$V_0/2 \pi ~ [\text{kHz}]$ & $2.8$ & $5.3$ & $7.9$ & $9.9$ & $11.2$ & $11.8$ & $12.1$
\end{tabular}
\caption{Realistic blockade radii $a_B$ and energies $V_0$ for Rb with $\omega_c \approx 2 \pi \cdot 130 \text{Hz}$ (i.e. $\ell_B \approx 1 \mu \text{m}$), $\Delta = 10 \Omega = 2 \pi \cdot 1 \text{GHz}$ as a function of the principle quantum number $n$.}
\label{tab:V0}
\end{table}
%%%%%%%%%%%%%%%%%%%%%%%%%%%%%%%%%%%%%%%%%%%%%%%%%%%%%%%%%%%%%%%%%%%%%%%%%%%%%%%%%%%%%%%
%%%%%%%%%%%%%%%%%%%%%%%%%%%%%%%%%%%%%%%%%%%%%%%%%%%%%%%%%%%%%%%%%%%%%%%%%%%%%%%%%%%%%%%
\begin{table}[b]
\centering
\begin{tabular}{c|
|p{1cm}|
%p{1cm}|
p{1cm}|
%p{1cm}|
p{1cm}|
%p{1cm}|
p{1cm}|
%p{1.2cm}|
%p{1.2cm}|
p{1.2cm}}
$\Delta [\text{MHz}] $ & 
 $1.0$ &
% $3.0$ &
$9.0$ &
%$30$ &
$80$ &
%$250$ &
$700$ &
%$2.0 \cdot 10^3$ &
% $6.5 \cdot 10^3$ &
$20 \cdot 10^3$ \\ \hline
$a_B ~ [\mu \text{m}]$ & 
$3.94$ &
% $3.28$ &
$2.73$ &
%$2.24$ &
$1.90$ &
%$1.57$ &
$1.32$ &
%$1.11$ &$0.91$ &
$0.76$ \\ \hline
$V_0/2 \pi ~ [\text{kHz}]$ & 
$0.0744$ &
%$0.210$ &
$0.574$ &
%$1.65$ &
$3.75$ &
%$9.37$ &
$20.8$ &
%$45.7$ &$107.8$ &
$240.0$
\end{tabular}
\caption{Realistic blockade radii $a_B$ and energies $V_0$ for Rb with $\omega_c \approx 2 \pi \cdot 130 \text{Hz}$ 
(i.e. $\ell_B \approx 1 \mu \text{m}$), $n = 46$ as a function of the detuning of the dressing laser $\Delta$. Note that 
the Rabi frequency $\Omega$ was chosen as $\Omega = 0.1 \Delta$.}
\label{tab:V0a}
\end{table}
%%%%%%%%%%%%%%%%%%%%%%%%%%%%%%%%%%%%%%%%%%%%%%%%%%%%%%%%%%%%%%%%%%%%%%%%%%%%%%%%%%%%%%%

One recognizes rather large values of $V_0$ up to hundreds of kHz, which can easily become comparable to or even exceed the typical LL splitting 
$\omega_c$. Thus different from all previously discussed interactions in cold gases the characteristic energy scales for Rydberg-Rydberg 
interactions must be limited by demanding LLL-approximation (i.e. $V_0 < \hbar \omega_c$). 

Table \ref{tab:Vms} lists some explicit values of higher pseudopotentials $V_m$ for the vdW interaction \eqref{eq:effPot}. They are also plotted 
in fig.\ref{fig1_5} together with contour lines of the continuous function $V(m)=V_m$ for $m \in \mathbb{R}$ defined as in \eqref{eq:defVm} above.
One recognizes that at given $a_B$ the first $V_m$ are \emph{approximately equal} until the radial extend $R_m=\sqrt{2m}\ell_B$ of the wavefunction in the relative coordinate is $\sim a_B$. The subsequent pseudopotentials decrease quickly. 

The relevance of the pseudopotentials becomes apparent if one considers the (bosonic analogue of the)
Laughlin wavefunction for the ground state of $N$ particles at filling $\nu=1/n$, with $n$ being an (even) integer. Denoting the coordinate of the $j'$th particle in the two-dimensional plane
by the normalized complex variable $z_j=(x_j + i y_j)/\ell_B$ the LN wave function reads \cite{LAUGHLIN1983}
\begin{equation}
 \psi_{\nu = 1/n}\bigl(z_1,\dots,z_N\bigr) \, \sim\, \prod_{i<j} \bigl(z_i-z_j\bigr)^n\, 
e^{-\frac{1}{4}\sum_k |z_k|^2}. 
\end{equation}
In the absence of interactions and any confinement potential all states are degenerate
and have zero energy. When considering the contribution of the interaction Hamiltonian to the
energy of the LN states one recognizes that the Jastow factors $\prod_{i<j} \left(z_i-z_j\right)^n$ 
eliminate furthermore all contributions from pseudopotentials $V_m$ with $m=0,2,\dots,n-2$.  
As a consequence the energy scale of the ground state at filling $\nu=1/m$ is set by the
largest unscreened pseudopotential, i.e. by $V_m$. 

For this reason the curve $\sqrt{2m}\, \ell_B = a_B$ separating the two regions in the contour plot of pseudopotentials fig.\ref{fig1_5} also separates two parameter regions with qualitatively different
physics for a given filling fraction $\nu$ 
\begin{equation}
 \sqrt{2/\nu}\, \ell_B < a_B,\quad \text{and}\quad \sqrt{2/\nu}\, \ell_B > a_B
\end{equation}
The first region corresponds to at most one particle per blockade area on average, while the second
corresponds to more than one particle in that area.

\onecolumngrid
%%%%%%%%%%%%%%%%%%%%%%%%%%%%%%%%%%%%%%%%%%%%%%%%%%%%%%%%%%%%%%%%%%%%%%%%%%%%%%%%%%%%%%%
\begin{center}
\begin{table}[t]
\centering
\begin{tabular}{|c||c||c|c|c|c|c|c|c|c|c|}
\hline
$a_B / \ell_B$& $V_0 [\tilde{C}_6 / \ell_B^6]$ & $V_2/V_0$ & $V_4/V_0$ & $V_6/V_0$ & $V_8/V_0$ & $V_{10}/V_0$ & $V_{12}/V_0$ & $V_{14}/V_0$ \\% & $V_{18}$ & $V_{20}$\\
\hline
$0.1 $&$3.015\cdot 10^{3}$ & $1.40\cdot 10^{-5} $ & $2.16\cdot 10^{-7} $ & $4.30\cdot 10^{-8} $ & $1.54\cdot 10^{-8} $ & $7.19\cdot 10^{-9} $ & $3.91\cdot 10^{-9} $ & $2.37\cdot 10^{-9} $ \\
$0.2 $&$1.871\cdot 10^{2}$ & $1.69\cdot 10^{-4} $ & $3.48\cdot 10^{-6} $ & $6.95\cdot 10^{-7} $ & $2.49\cdot 10^{-7} $ & $1.16\cdot 10^{-7} $ & $6.31\cdot 10^{-8} $ & $3.82\cdot 10^{-8} $ \\
$0.3 $&$3.652\cdot 10^{1}$ & $6.93\cdot 10^{-4} $ & $1.78\cdot 10^{-5} $ & $3.56\cdot 10^{-6} $ & $1.27\cdot 10^{-6} $ & $5.95\cdot 10^{-7} $ & $3.23\cdot 10^{-7} $ & $1.96\cdot 10^{-7} $ \\
$1 $ &$2.455\cdot 10^{-1}$ & $3.43\cdot 10^{-2} $ & $2.53\cdot 10^{-3} $ & $5.28\cdot 10^{-4} $ & $1.89\cdot 10^{-4} $ & $8.82\cdot 10^{-5} $ & $4.80\cdot 10^{-5} $ & $2.91\cdot 10^{-5} $ \\
$2 $ &$9.816\cdot 10^{-3}$ & $0.223$ & $4.56\cdot 10^{-2} $ & $1.24\cdot 10^{-2} $ & $4.66\cdot 10^{-3} $ & $2.20\cdot 10^{-3} $ & $1.20\cdot 10^{-3} $ & $7.28\cdot 10^{-4} $ \\
$3 $&$1.172\cdot 10^{-3}$ & $0.505$ & $0.203$ & $8.06\cdot 10^{-2} $ & $3.55\cdot 10^{-2} $ & $1.77\cdot 10^{-2} $ & $9.91\cdot 10^{-3} $ & $6.04\cdot 10^{-3} $ \\
$4 $&$2.314\cdot 10^{-4}$ & $0.740$ & $0.450$ & $0.250$ & $0.137$ & $7.75\cdot 10^{-2} $ & $4.63\cdot 10^{-2} $ & $2.92\cdot 10^{-2} $ \\
$5 $&$6.279\cdot 10^{-5}$ & $0.877$ & $0.680$ & $0.479$ & $0.322$ & $0.212$ & $0.140$ & $9.47\cdot 10^{-2} $ \\
$10 $ &$9.996\cdot 10^{-7}$& $0.996$ & $0.987$ & $0.970$ & $0.944$ & $0.908 $ & $0.863$ & $0.811$ \\ \hline
\end{tabular}
\caption{Leading order bosonic ($m$ even) pseudopotentials $V_m$ for the potential eq.\eqref{eq:effPot} discussed in the main text.}
\label{tab:Vms}
\end{table}
\end{center}
%%%%%%%%%%%%%%%%%%%%%%%%%%%%%%%%%%%%%%%%%%%%%%%%%%%%%%%%%%%%%%%%%%%%%%%%%%%%%%%%%%%%%%%
\twocolumngrid

%%%%%%%%%%%%%%%%%%%%%%%%%%%%%%%%%%%%%%%%%%%%%%%%%%%%%%%%%%%%%%%%%%%%%%%%%%%%%%%%%%%%%
%%%%%%%%%%%%%%%%%%%%%%%%%%%%%%%%%%%%%%%%%%%%%%%%%%%%%%%%%%%%%%%%%%%%%%%%%%%%%%%%%%%%%
\section{ground state for small blockade radii}
%%%%%%%%%%%%%%%%%%%%%%%%%%%%%%%%%%%%%%%%%%%%%%%%%%%%%%%%%%%%%%%%%%%%%%%%%%%%%%%%%%%%%
%%%%%%%%%%%%%%%%%%%%%%%%%%%%%%%%%%%%%%%%%%%%%%%%%%%%%%%%%%%%%%%%%%%%%%%%%%%%%%%%%%%%%

In the following we discuss the ground state of the system in
the case of a small blockade radius $a_B<\ell_B \sqrt{2/\nu}$, where its effect can be disregarded.

%%%%%%%%%%%%%%%%%%%%%%%%%%%%%%%%%%%%%%%%%%%%%%%%%%%%%%%%%%%%%
\subsection{The $\nu=1/2$ and $\nu=1/4$ ground states.}
%%%%%%%%%%%%%%%%%%%%%%%%%%%%%%%%%%%%%%%%%%%%%%%%%%%%%%%%%%%%%

For $a_B/\ell_B \rightarrow 0$ the first two pseudopotentials diverge
\begin{equation}
 V_0 \approx \frac{3}{8} \frac{\tilde{C}_6}{\ell_B^6} \left(\frac{\ell_B}{a_B}\right)^{4}, \enspace V_2 \approx \frac{1}{2^6} \l 0.571 - 
\ln \frac{a_B}{\ell_B} \r \frac{\tilde{C}_6}{\ell_B^6},
\label{eq:VmScaling}
\end{equation}
while all $V_{m> 2}$ converge. Therefore, the $\nu=1/2,1/4$ bosonic LN states are exact ground states.
To understand the effect of finite (but small) blockade radii we performed exact diagonalization (ED) calculations in spherical geometry for small particle numbers $N$. Table \ref{tab:ovlps} lists overlaps to the Laughlin states for $N=6$ particles and several $a_B$. The results remain valid even for larger particle numbers: We obtain overlaps squared of e.g. $0.992$ for $\nu=1/2$ and $N=10$ and $0.974$ for $\nu=1/4$ and $N=7$.
Some remarks on the ED simulations are given in the Appendix.

%%%%%%%%%%%%%%%%%%%%%%%%%%%%%%%%%%%%%%%%%%%%%%%%%%%%%%%%%%%%%%%%%%%%%%%%%%%%%%%%%%%%%%%
\begin{table}[h]
\centering
\begin{tabular}{|c||c|c|c|c|}
\hline
 $a_B/ \ell_B$ & $\nu=1/2$ & $\nu=1/4$ \\ \hline
 $6.1$ &$1.690\cdot 10^{-09}$ & $0$  \\ \hline
$5.1$ &$4.935\cdot 10^{-10}$ & $0$ \\ \hline 
$4.1$ &$7.303\cdot 10^{-10}$ & $0$ \\ \hline 
$3.1$ &$2.594\cdot 10^{-03}$ & $0.9026$ \\ \hline
$2.1$ &$0.9940$ & $0.9696$  \\ \hline
$1.1$ &$0.9998$ & $0.9960$ \\ \hline
$0.1$ & $0.999999999978$ & $0.99988$ \\ \hline
\end{tabular}
\caption{Overlaps squared of the ground state to the Laughlin states at fillings $\nu=1/2,1/4$. They were obtained using 
ED at $N=6$ in spherical geometry. We write $0$ when our value is below the numerical precision, i.e. $< 10^{-15}$.}
\label{tab:ovlps}
\end{table}
%%%%%%%%%%%%%%%%%%%%%%%%%%%%%%%%%%%%%%%%%%%%%%%%%%%%%%%%%%%%%%%%%%%%%%%%%%%%%%%%%%%%%%%

%%%%%%%%%%%%%%%%%%%%%%%%%%%%%%%%%%%%%%%%%%%%%%%%%%%%%%%%%%%%%%%%%%%%%
\subsection{Ground states at small fillings}
%%%%%%%%%%%%%%%%%%%%%%%%%%%%%%%%%%%%%%%%%%%%%%%%%%%%%%%%%%%%%%%%%%%%%

Now we want to address the physics for small fillings, i.e. $\nu<1/4$, again for $a_B \lesssim \ell_B$. Due to the non-local interaction
a natural ground state candidate in this regime is the noncorrelated Wigner crystal (NWC) \cite{MAKI1983} described by the wavefunction
\begin{equation}
\psi_{\text{NWC}}(z) = \mathcal{S} ~ \prod_j e^{- \l |z_j-R_j|^2 + z_j R_j^* - z_j^* R_j \r/4},
\label{eq:nonCorrWC}
\end{equation}
where $R_j \in \mathbb{C}$ defines a lattice in the plane of complex coordinates $z_j=(x_j+i y_j)/\ell_B$ and $\mathcal{S}$ stands for complete symmetrization. 

%***********************************************
\subsubsection*{variational ground state energy}
%***********************************************

\begin{figure}[b]
\begin{center}
\epsfig{file=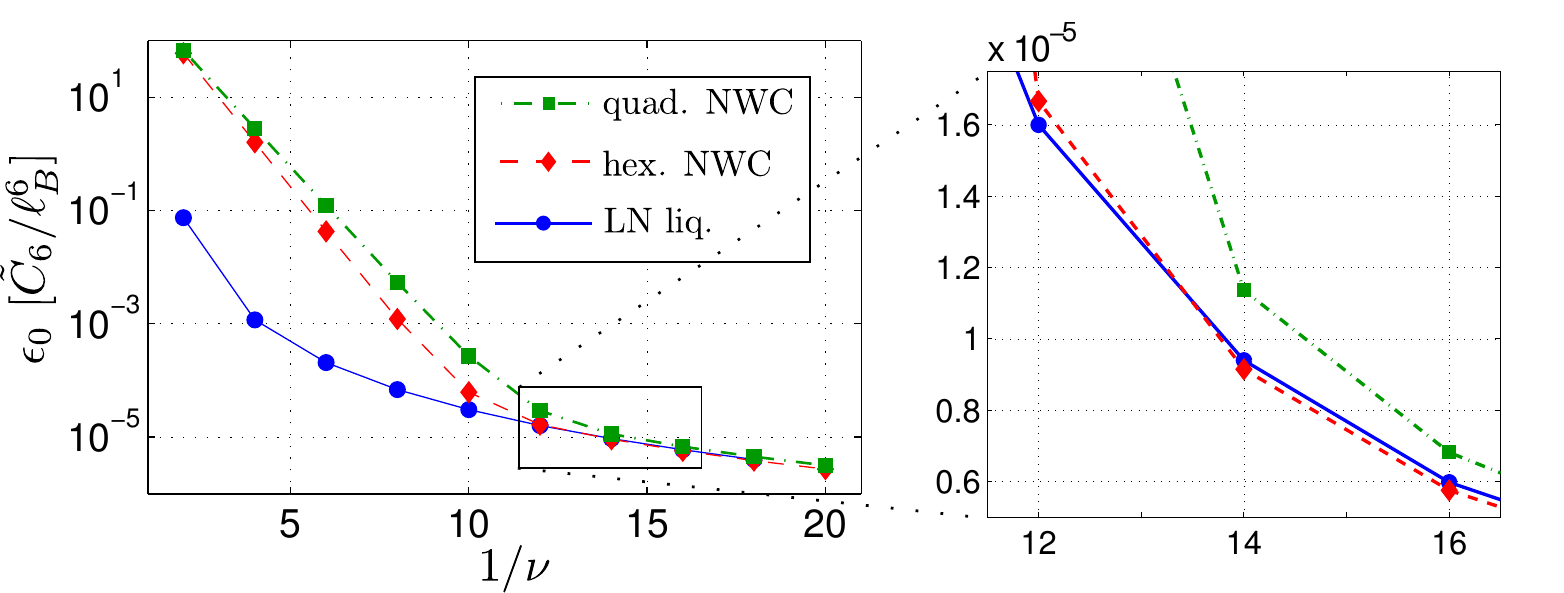, width=0.5\textwidth}
\end{center}
\caption{(color online). Comparison of the variational energies per particle $\epsilon_0$ (at fillings $\nu=1/2, 1/4, 1/6,...$ corresponding to the bosonic Laughlin states) 
for the different trial wavefunctions at $a_B=0.2 \ell_B$ (LN - Laughlin states, NWC - noncorrelated Wigner crystal of bare particles.)}
\label{fig:variational-energies}
\end{figure}

To see if the NWC could be a ground state we compare the corresponding variational energy per particle $\epsilon_{\text{NWC}}$ to that of the LN state.
To this end we generalize the expression for $\epsilon_\text{NWC}$ derived in \cite{MAKI1983} to bosons:
\begin{equation}
 \epsilon_{\text{NWC}} = \frac{1}{2} \frac{\tilde{C}_6}{\ell_B^6} \sum_{j \neq 0} 
\frac{e^{-|R_j|^2/4}}{1 + 
\frac{1}{2} e^{-|R_j|^2/2}} \cdot K(\xi=|R_j|),
\label{eq:nonCorrWCenergy}
\end{equation}
\begin{equation*}
 K(\xi)=\int_0^{\infty}\!\! \frac{dr ~ r e^{-r^2/4}}{r^6 + \l a_B/\ell_B \r^6} \l I_0 \l \frac{r 
\xi}{2} \r + J_0 \l \frac{r \xi}{2} \r \r,
\end{equation*}
where $I_0,J_0$ denote Bessel functions. The integral can be handled numerically and the lattice sum can easily be performed.
Like for electrons \cite{LAM1984,MAKI1983} and dipolar fermions \cite{Baranov2008}, the lattice that minimizes the NWC energy is a hexagonal one, see fig.\ref{fig:variational-energies}. 
%%%%%%%%%%%%%%%%%%%%%%%%%%%%%%%%%%%%%%%%%%%%%%%
%%%%%%%%%%%%%%%%%%%%%%%%%%%%%%%%%%%%%%%%%%%%%%%%%

In fig.\ref{fig:variational-energies} the energies of different NWCs and the LN liquid are plotted for small $a_B=0.2 \ell_B$ and even values of $1/\nu$. The energy of the LN liquid was calculated using the 
standard plasma analogy \cite{LAUGHLIN1983} for $N=100$ particles for all examined $\nu$ using a Metropolis Monte Carlo (MC) algorithm \cite{METROPOLIS1953}. We find a transition from LN to 
a NWC for 
\begin{equation}
 \nu\le\nu_\text{NWC}=\frac{1}{14},
\end{equation}
although for $\nu=1/12$ the energy difference is so small that we can not exclude a transition here. 
Interestingly this value of the critical filling is far below the corresponding value of $\nu\approx 1/7$ found for pure Coulomb and dipolar interacting fermions \cite{LAM1984,Baranov2008,Pan2002}. 
Due to the more local nature of the vdW interaction potential, this is not surprising: For point-like interactions there is no crystallization at all.

%**********************************************************
\subsubsection*{stability of Wigner-crystal}
%**********************************************************

In an alternative approach 
we can determine the stability region of a lattice using e.g. the 
Lindemann criterion. To this end we calculate the 
Lindemann parameter $\gamma\equiv \sqrt{\left\langle \delta \textbf{u}^2 \right\rangle} / a$ from
the phonon spectrum in harmonic and nearest-neighbor approximation following \cite{FUKUYAMA1976}, where 
$a$ denotes the (hexagonal) lattice constant and $\delta \textbf{u}$ the atomic displacement from the lattice $\left\lbrace R_j \right \rbrace$. 
The phenomenological criterion states that the crystal melts when $\gamma$ exceeds a critical value 
$\gamma_c$. In our case $
 \gamma = 0.57 \sqrt{\nu}$
for $a_B < \ell_B$ at zero temperature, and $\gamma_c\approx 0.28$ \cite{LOZOVIK1985,Baranov2008}. Thus 
the crystal is expected to be stable for 
\begin{equation}
 \nu < \nu_{\rm Ld} = \frac{1}{4},
\end{equation}
similar to the result found in dipolar 
systems \cite{Baranov2005,Baranov2008}.
This value differs significantly from the transition point to a NWC, $\nu_{\text{NWC}}=1/14$, found 
above!

%*********************************************************
\subsubsection*{gap to collective excitations and spatial correlations}
%*********************************************************

\begin{figure}[t]
\begin{center}
\epsfig{file=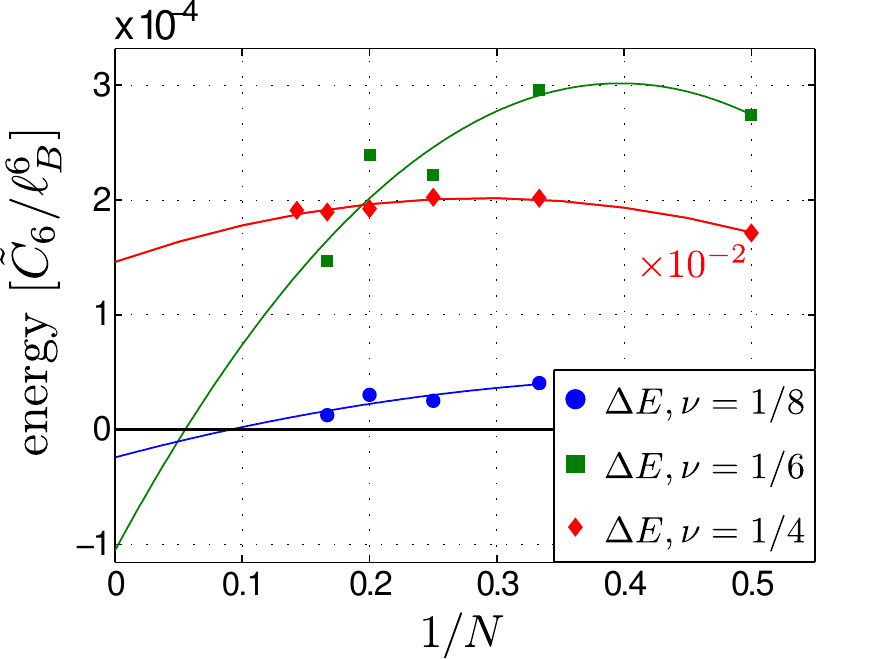,width=0.6\columnwidth}
\end{center}
\caption{(color online). Gap to collective excitations $\Delta E$ at $\nu=1/4,1/6,1/8$ and for $a_B=0.2 \ell_B$. Note that for $\nu=1/4$ energies were scaled down by a factor of $100$. ED in spherical geometry was used and finite-size corrections were performed as described in the Appendix. Solid lines: quadratic fits.}
\label{figSupp1a}
\end{figure}

For small system sizes (see appendix) and small $a_B \lesssim \ell_B$ the exact ground state 
has relatively large overlap to the LN liquid and the spectrum shows a low-lying exciton branch.
On the other hand the analysis of the previous section suggests that in the thermodynamic limit
the true ground state may not be a LN liquid. Thus to investigate the nature of the ground 
state in the region of intermediate fillings we calculate the energy gap $\Delta E$ of 
LN states to collective excitations using ED in spherical geometry. 
More specifically $\Delta E$ is the difference from the ground to first excited state, where the 
latter can well be described as a density wave and exhibits a roton minimum \cite{GIRVIN1986,Kamilla1996}.
In Fig.\ref{figSupp1a} $\Delta E$ is plotted for $\nu=1/6, 1/8$ for finite size systems and extrapolations to the thermodynamic limit are shown. Also shown are the results for $\nu=1/4$ for comparison, which clearly yield a positive gap in the thermodynamic limit. Although the system sizes which we were able to reach with our ED are not very large they do allow for an extrapolation to the infinite size limit and indicate negative gaps - i.e. instability - for $\nu=1/6$ and $\nu=1/8$.

\begin{figure}[b]
\begin{center}
\epsfig{file=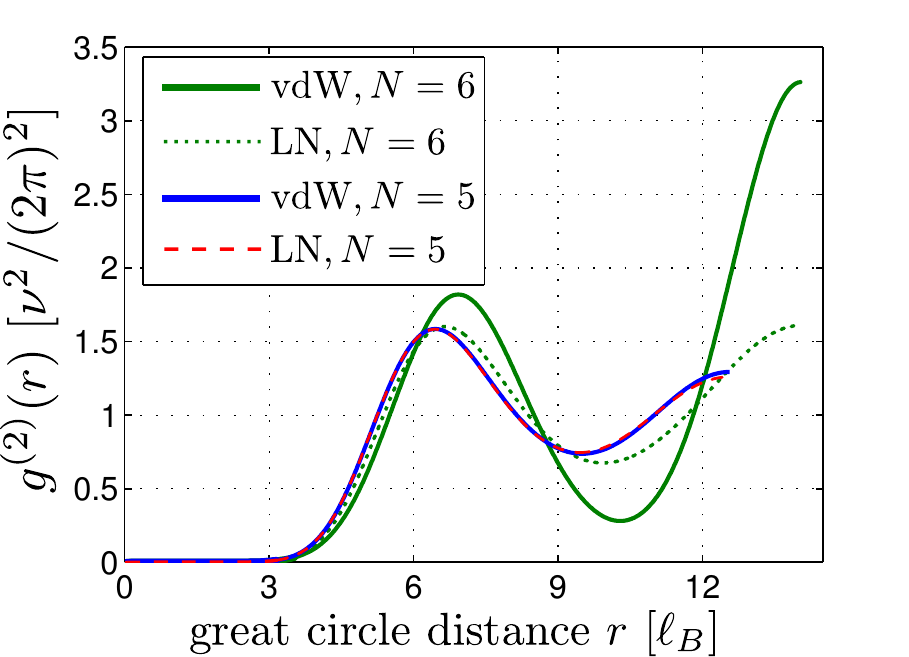,width=0.6\columnwidth}
\end{center}
\caption{(color online). Second order correlation functions, $g^{(2)}(r)$, of the ground state for 
van-der-Waals (vdW) interactions at $a_B=0.2 \ell_B$ and $\nu=1/8$, compared to those of the LN 
liquids at the corresponding system size. ED in spherical geometry was used. }
\label{figSupp1b}
\end{figure}

In fig. \ref{figSupp1b} second order correlations, $g^{(2)}(r) = \langle \hat{\psi}^\dagger(0)\hat{\psi}^\dagger(r) \hat{\psi}(r) \hat{\psi}(0)\rangle$ are shown for $\nu=1/8$ and $a_B=0.2 \ell_B$. We find strongly enhanced oscillations for $N=6$ signalizing a trend towards long-range order. For $N=5$ this effect is absent, which can be attributed to a more general finite-size effect: Already for the gaps (fig. \ref{figSupp1a}) we observe a slightly different behavior for $N$ even/odd and $\nu=1/6,1/8$. For $N=2,4,6$, $\Delta E$ is smaller than for $N=3,5$ and we find the same behavior for the ground state energy. This is another indication for crystallization, since also for a classical vdW crystal on a sphere the ground state energy per particle shows these oscillations
due to incommensurability.

These numerical findings for small systems support the result of the Lindemann stability analysis, that crystallization occurs below $\nu=1/4$. That raises the question about the nature of the ground state for fractional fillings in the range $1/4 > \nu \geq 1/14$.

%%%%%%%%%%%%%%%%%%%%%%%%%%%%%%%%%%%%%%%%%%%%%%%%%%%%%
%%%%%%%%%%%%%%%%%%%%%%%%%%%%%%%%%%%%%%%%%%%%%%%%%%%%%

%%%%%%%%%%%%%%%%%%%%%%%%%%%%%%%%%%%%%%%%%%%%%%%%%%%%%
%%%%%%%%%%%%%%%%%%%%%%%%%%%%%%%%%%%%%%%%%%%%%%%%%%%%%

%%%%%%%%%%%%%%%%%%%%%%%%%%%%%%%%%%%%%%%%%%%%%%%%%%%%%%%%%%%%%%%%%%%%%%%%%%%
\subsection{Correlated Wigner crystal of composite particle}
%%%%%%%%%%%%%%%%%%%%%%%%%%%%%%%%%%%%%%%%%%%%%%%%%%%%%%%%%%%%%%%%%%%%%%%%%%%%

Instead of a NWC the ground state in the region $\nu_\text{Ld} > \nu\geq\nu_\text{NWC}$ could be a crystal of 
composite particles (CPs) \cite{JAIN1989}, termed \emph{correlated} Wigner crystal (CWC) \cite{Yi1998,Chang2005a}.  
As for the LN states the 
first pseudopotentials are screened for CPs and only the effects of the long-range tails of the 
repulsive interaction potential remain. As the latter should be the same for CPs as for bare particles the stability arguments given
above remain valid and a CWC should form already for $\nu < \nu_{\text{Ld}}=1/4$.

\begin{figure}
\begin{center}
\epsfig{file=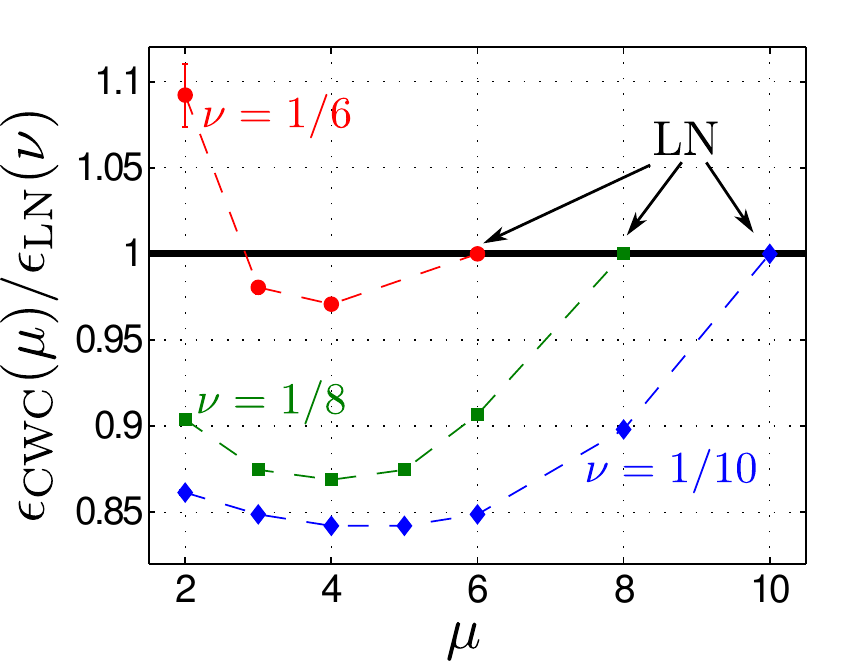, width=0.6\columnwidth}
\end{center}
\caption{(color online) Variational energy per particle of the hexagonal CWC of $\CP{\mu}$, $\epsilon_{\text{CWC}}(\mu)$, in units of the LN state energy $\epsilon_{\LN}(\nu)$ at filling $\nu$. Results were obtained in Metropolis MC simulations with $N=91$ particles.
}
\label{fig:CWC-variational-energies}
\end{figure}

CWC variational wavefunctions of the form 
\begin{equation*}
  \psi_{\text{CWC}}^{(\mu)}(z) = \mathcal{P}_{\LLL} \Ys^{\mu} \psi_{\text{NWC}}(z)
\end{equation*}
were investigated in \cite{Yi1998} using MC simulations based on the plasma analogy, and the critical filling $\nu \approx 1/7$ in electronic, i.e. fermionic systems was correctly predicted. Here $\mathcal{P}_{\LLL}$ is a projector to the LLL and $(z_j-z_k)^\mu$ are Jastrow factors describing the formation of $\mu$- composite bosons $\CB{\mu}$ (fermions $\CF{\mu}$) for $\mu$ even (odd). 
We here performed similar MC simulations 
and calculated the variational energy of hexagonal CWCs for different $\mu$.
From the plasma analogy it can be concluded \cite{Yi1998} that the lattice constant $a$ used for the NWC have to be rescaled in the definition of the CWC wavefunction in order to produce the desired density at filling $\nu$: 
\begin{equation}
 a \rightarrow a \l 1- \mu \cdot \nu \r.
\end{equation}
In the MC simulations we take into account only direct but no exchange terms but have checked that the first exchange corrections are negligible.
Our results are shown in fig.\ref{fig:CWC-variational-energies} and we observe that for $\nu \leq 1/6$ CWCs are energetically favorable compared to LN states. 

%%%%%%%%%%%%%%%%%%%%%%%%%%%%%%%%%%%%%%%%%%%%%%%%%%%
%%%%%%%%%%%%%%%%%%%%%%%%%%%%%%%%%%%%%%%%%%%%%%%%%%%%%

We thus expect crystalline order already for $\nu < \nu_\text{Ld}=1/4$ in agreement with the stability analysis and long before the simple NWC becomes energetically favorable at $\nu =\nu_\text{NWC}$.
More specifically we expect a $\CB{4}$ CWC ground state at $\nu=1/6$ and $\nu=1/8$. For $\nu=1/10$ the $\CB{4}$ and $\CF{5}$ CWCs are equal in energy within our MC errors and we can not make a final conclusion about the underlying CPs. Most importantly the variational energy of the composite
crystals is for a larger range of fractional fillings substantially below the value of the
non-correlated Wigner-crystal.

That crystals of composite particles are formed for $\nu < 1/4$ is further supported by the ED results for the second-order correlation function plotted in fig.\ref{figSupp1b}.
For larger distances $r$ one recognizes enhanced oscillations of $g^{(2)}(r)$ as compared to those expected for LN liquids. On the other hand, the correlations decay still according to a power law
for small $r$, e.g. $g^{(2)}(r)\sim r^{11}$ for $N=6$ at $\nu=1/8$. This is a direct indication that the ground state physics of these systems is dominated by $\CP{\mu}$s with $\mu$ between $4$ and $6$, as predicted above.

%%%%%%%%%%%%%%%%%%%%%%%%%%%%%%%%%%%%%%%%%%%%%%%%%%%%%%%%%%%%%%%%%%%%%%%%%%%%%%%%%%
%%%%%%%%%%%%%%%%%%%%%%%%%%%%%%%%%%%%%%%%%%%%%%%%%%%%%%%%%%%%%%%%%%%%%%%%%%%%%%%%%%
\section{Effects of finite blockade radius}
%%%%%%%%%%%%%%%%%%%%%%%%%%%%%%%%%%%%%%%%%%%%%%%%%%%%%%%%%%%%%%%%%%%%%%%%%%%%%%%%%%
%%%%%%%%%%%%%%%%%%%%%%%%%%%%%%%%%%%%%%%%%%%%%%%%%%%%%%%%%%%%%%%%%%%%%%%%%%%%%%%%%%

We now discuss the effects of the saturation of the effective
vdW interaction potential, eq.(\ref{eq:effPot}), for distances less than the blockade radius $a_B$. 
As we will show the competition of the magnetic length $\ell_B$  with the new length $a_B$
will give rise to new physics which has similarities with physics of higher LLs in the solid
state context.
Specifically we consider the case where there are more than two particles per blockade 
area $A_B=\pi a_B^2$, i.e. the regime
$a_B\gtrsim \sqrt{2/\nu} \ell_B$.

%%%%%%%%%%%%%%%%%%%%%%%%%%%%%%%%%%%%%%%%%%%%%%%%%%%%%%%%%%%%%%%%%%%%%%%%%%%%%%%%%%
\subsection{Bubble crystal at small fillings.}
%%%%%%%%%%%%%%%%%%%%%%%%%%%%%%%%%%%%%%%%%%%%%%%%%%%%%%%%%%%%%%%%%%%%%%%%%%%%%%%%%%

\begin{figure}[t]
\begin{center}
\epsfig{file=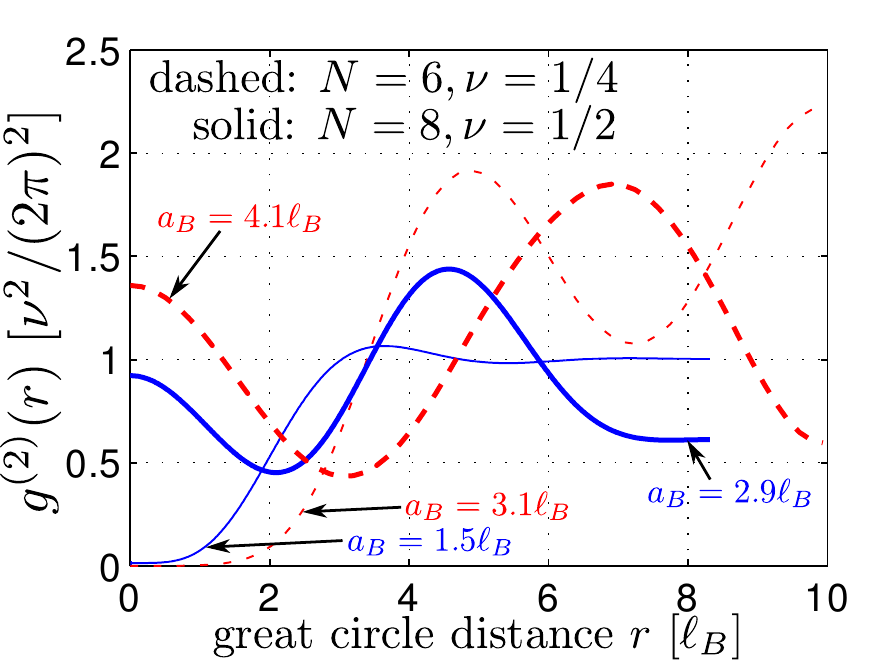, width=0.65\columnwidth}
\end{center}
\caption{(color online). Second order correlation function for the ground state
for increasing values of $a_B$ for $N=8$, solid ($N=6$, dashed) for $\nu=1/2$ ($\nu =1/4$). For
large $a_B$ symmetry is broken ($L\ne 0$ thick lines), in contrast to small $a_B$
($L=0$ thin lines).
}
\label{fig:FQHE-finite-ab}
\end{figure}

ED results obtained on spheres suitable for LN states show that for $a_B \lesssim 
\sqrt{2/\nu}\ell_B$ and $\nu =1/4$ or $1/2$, 
the ground state is a $\nu$ LN state with vanishing angular momentum $L=0$. For $a_B \approx \sqrt{4/\nu} \ell_B$ a transition to a $L\neq0$ state is observed, which after mapping from sphere (radius $R$) to plane (momentum $k$), $kR=L$ \cite{Haldane1990}, corresponds to a breaking of translational invariance. Fig.\ref{fig:FQHE-finite-ab} shows numerical results for the two-particle correlation $g^{(2)}(r)$ for different ratios of $a_B/\ell_B$. 
For small $a_B$ ($L=0$ phase) we find LN-like correlations while for large $a_B$ ($L \neq 0$ phase) we find particle bunching at $r=0$, which indicates clustering. Clustering of $k$ particles can also be seen in the $k+1$st-order density correlations,
\begin{equation*}
 g^{(k+1)}(z) \equiv \Bigl\langle  \left(\hat{\Psi}^{\dagger}(0)\right)^{k} 
\hat{\Psi}^{\dagger}(z)\hat{\Psi}(z)\left(\hat{\Psi}(0)\right)^{k}  \Bigr\rangle.
\end{equation*}
E.g. for $a_B=2.9 \ell_B$, $k=2$ particles cluster, resulting in a very small
$g^{(3)}(0)\approx 2\times 10^{-5} \l \nu/2 \pi \r^3$ (for $\nu=1/2, N=8$), while $g^{(2)}(0)=0.9 \l \nu / 2 \pi \r^2$ is still large. 

%%%%%%%%%%%%%%%%%%%%%%%%%%%%%%%%%%%%%%%%%%%%%%%%%%%%%
%%%%%%%%%%%%%%%%%%%%%%%%%%%%%%%%%%%%%%%%%%%%%%%%%%%%%

These numerical findings can be explained as follows: For $a_B > \sqrt{4/\nu}\ell_B$ where there are more than two particles per $A_B$, the long-range 
contribution to the interaction energy can be reduced by bringing the particles inside $A_B$ together.
At the same time there is no energy penalty from interactions inside $A_B$ since the potential (\ref{eq:effPot})  is constant for $r \lesssim a_B$. This provides a pairing mechanism.
Assuming that all particles within $A_B$ undergo clustering, one expects a transition from $k-1$ to $k$ particles per cluster at $a_B^{(k)}= \sqrt{2 k/\nu} \cdot \ell_B$.

\begin{figure}
\begin{center}
\epsfig{file=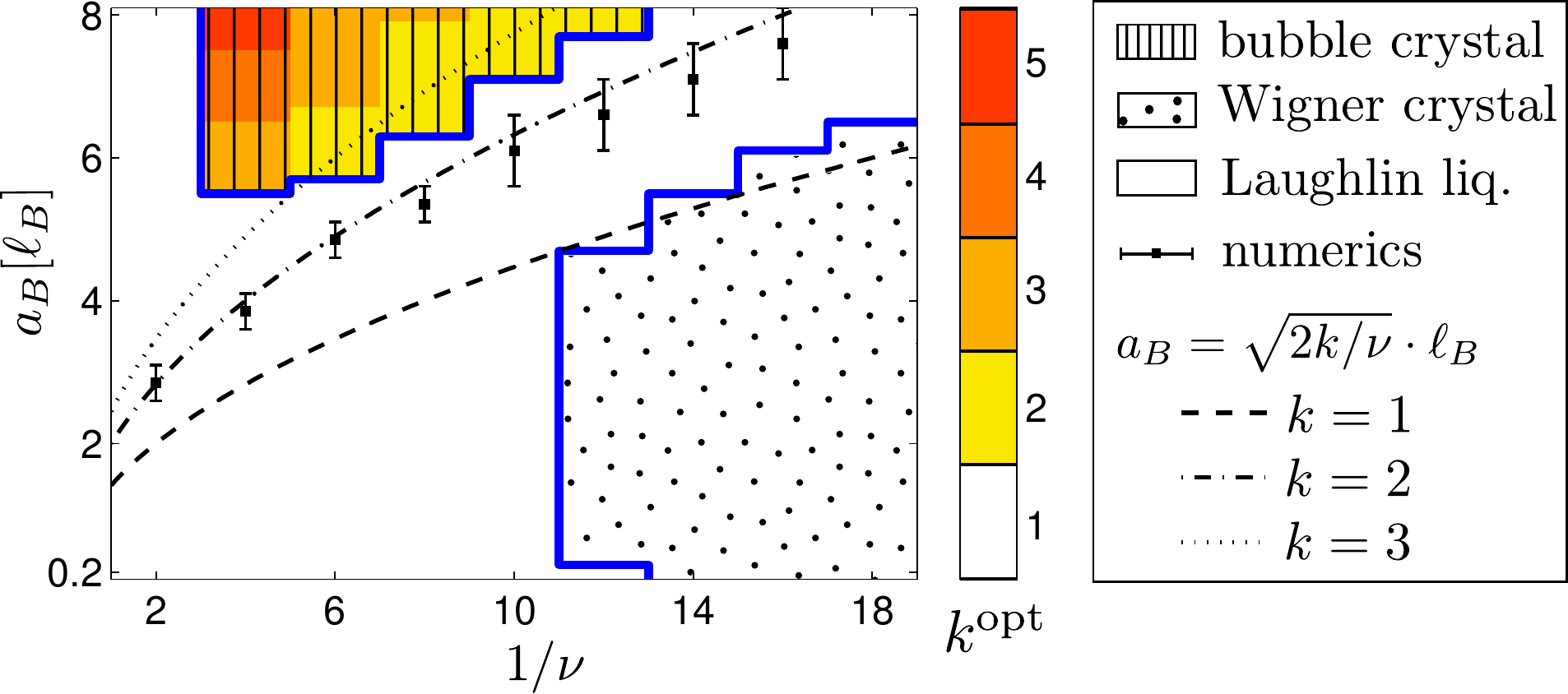, width=0.45\textwidth}
\end{center}
\caption{(color online). Phase diagram (for $1/\nu \in 2\mathbb{N}$) obtained from the comparison of variational energies of LN state ($N=100$ using Metropolis MC) and bubble crystal. The optimal number $k^{\text{opt}}$ of particles per cluster (color code) was used in the BC region. The squares with error bars show the phase transition from LN to a symmetry-breaking state obtained numerically (ED, spherical geometry).}
\label{fig5}
\end{figure}

By analogy we expect the formation of a NWC consisting of clusters, when the reduced filling of the system with clusters $\nu_{\cl}= \nu/ k$ is small. This state is referred to as bubble crystal (BC) and was 
e.g. considered for electrons in a weak magnetic field where $\nu >1$, i.e. beyond the LLL \cite{Koulakov1996}. The present 
situation is comparable since the effective interaction in a higher LL is ``smeared out'' around $r=0$, making it qualitatively similar to our interaction eq. \eqref{eq:effPot}.
A BC phase was also predicted for dipolar bosons at filling $\nu=1/2$ with large finite-thickness effects \cite{Cooper2005}.
To verify that BCs are indeed good ground state candidates for large blockade radii we calculate the variational
energy per particle by generalizing \eqref{eq:nonCorrWCenergy}. This yields 
\begin{equation}
  \epsilon_{\text{BC}}(k;\nu) = k \cdot \epsilon_{\text{NWC}}\l \nu=\nu_{\cl}\r + \l k -1 \r V_0,
\label{eq:epsClWC}
\end{equation}
where the second term describes the binding energy per particle required for the cluster formation. In fig.\ref{fig5} the ground state phase diagram is shown 
determined by comparing the variational energies of the LN liquid, the NWC and the BC. 
Also shown are the transition points between symmetry conserving ($L=0$) and symmetry breaking ($L\neq 0$) states obtained from ED. One recognizes that 
BC ground states with $k\geq2$ particles per cluster exist for $\nu \leq 1/4$ and $a_B \gtrsim a_B^{(2)}$.

%%%%%%%%%%%%%%%%%%%%%%%%%%%%%%%%%%%%%%%%%%%%%%%%%%%%%
%%%%%%%%%%%%%%%%%%%%%%%%%%%%%%%%%%%%%%%%%%%%%%%%%%%%%

%%%%%%%%%%%%%%%%%%%%%%%%%%%%%%%%%%%%%%%%%%%%%%%%%%%%%%%%%%%%%%%%
\subsection{Speculations for large filling: Cluster liquids}
%%%%%%%%%%%%%%%%%%%%%%%%%%%%%%%%%%%%%%%%%%%%%%%%%%%%%%%%%%%%%%%%

For $\nu=1/2$ the BC is higher in energy than the LN state for {\it all} $a_B$ and for sufficiently
large $a_B \gtrsim a_B^{(2)}$ the LN state is no longer a good ground state candidate. Therefore
the nature of the ground state for large $a_B$ and large filling is an open question.
In the following we will give some speculative arguments about the ground state in this
parameter region. In contrast to other regions here only numerical results 
from exact diagonalization for up to $N=10$ particles are available and thus 
it is difficult to make definite statements. Clearly further studies are needed here.
Yet our preliminary studies show that this parameter regime could be a very interesting one.

Besides clustered crystalline states of the type discussed above, many different correlated cluster {\em liquids} have been proposed and discussed in the context of the FQHE in higher LLs. Most prominent are the Read-Rezayi (RR) $\mathbb{Z}_k$ parafermion states \cite{Read1999} which have recently been generalized using totally symmetric Jack polynomials \cite{Bernevig2008}. The Haffnian (Hf) \cite{Green2001} is another example not contained in the Jack series. These states are characterized by the number of particles per cluster $k$ (e.g. $k=2$ for Hf) and the exponent $r$ which determines the short-distance behavior of the correlations $g^{(k+1)}(z)\sim |z|^{2 r}$.
For Hf $r=4$ and for all RR states $r=2$. For filling $\nu = 1$ the Moore-Read Pfaffian (Pf) \cite{MOORE1991} (the $k=2$ RR state) was shown to be the ground state for dipolar \cite{Rezayi2005} and contact \cite{Regnault2003} interactions corresponding to $a_B \ll \ell_B$. 
We also find a large overlap to the Pf of e.g. $\left| \left\langle \psi(N=12) | \text{Pf} \right\rangle \right|^2=0.90$ for $a_B=\ell_B$. Since the potential \eqref{eq:effPot} provides a clustering mechanism also for $\nu<1$, the cluster liquids are reasonable trial wavefunctions for our situation as well. 

%%%%%%%%%%%%%%%%%%%%%%%%%%%%%%%%%%%%%
\begin{figure}[t]%[hbt]
\begin{minipage}{0.5\textwidth}
 \epsfig{file=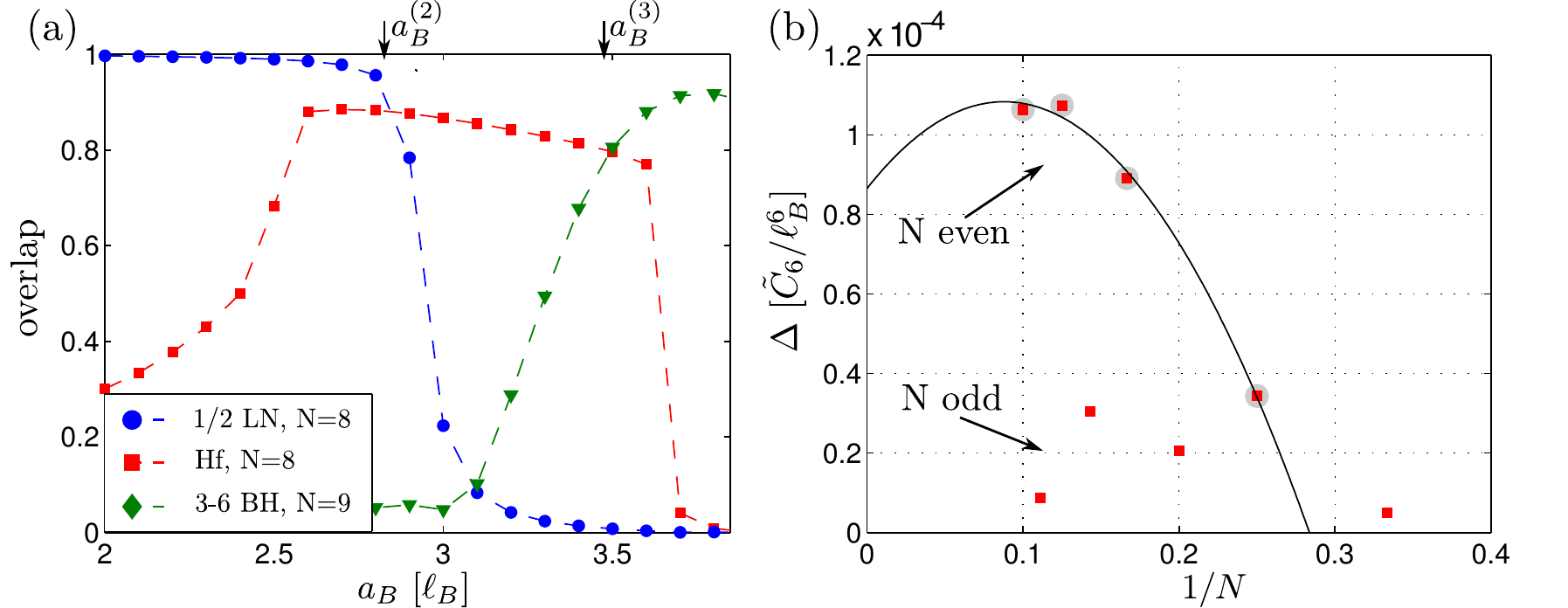, width=\textwidth}
\end{minipage}
\caption{(color online). ED results in spherical geometry. (a) Overlap of the numerically determined ground states to the different trial wavefunctions. (b) Finite size approximation of the gap to collective excitations for the ground state at the system size of the Hf (at $a_B=3 \ell_B$). Finite size corrections were carried out following \cite{DAMBRUMENIL1989}. The different behavior of $\Delta$ for $N$ even/ odd is another indication of cluster formation with $k=2$ particles per cluster.}
\label{fig6}
\end{figure}
%%%%%%%%%%%%%%%%%%%%%%%%%%%%%%%%%%%%%%%%

In the following we will focus on filling $\nu=1/2$, where the bubble crystal (BC) is no ground state. 
% candidate. 
We investigate the three simplest $\nu=1/2$ cluster liquids, namely the LN state ($k=1,r=2$), the Haffnian ($k=2,r=4$) and the $k=3, r=6$ Jack polynomial. We will refer to the latter as 3-6 BH (Bernevig, Haldane) state. As in the BC case, we expect a transition from $k-1$ to $k$ particles per cluster at $a_B^{(k)}$. This estimates the transition from LN to Hf to be at $a_B^{(2)}\simeq 2.8 \ell_B$ and from Hf to 3-6 BH at $a_B^{(3)}\simeq 3.5 \ell_B$. In fig.\ref{fig6} (a) we show the numerically obtained overlaps of the ground state wavefunctions with the trials. (The 3-6 BH state is obtained from the ``Jack generator'' \cite{Bernevig2009a}.) Note that the calculations were done for spheres of the \emph{different} sizes supporting the respective trial ground states (see supplementary). We find that for every $a_B$ at least one of the overlaps takes a substantial value.
This finding must be taken with care however: It is known that when a trial wavefunction has a large overlap to the exact ground state for small systems, this does not imply a good description in the thermodynamic limit.
The transitions between the different trial states in fig.\ref{fig6} are reasonably well described 
by the above estimates. Coexistence of phases is a finite size effect, at least partly caused by the different system sizes. 

We close our discussion of cluster liquids by investigating the ground state excitation gap. In fig.\ref{fig6} (b) finite size approximations are shown for the Hf state which suggest that it is an incompressible ground state in the thermodynamic limit. This is a surprising result since the Hf is generally believed to describe compressible states. We note that the arguments given by Green \cite{Green2001} that the Hf lies on a phase transition and should be gapless rely on the fact that the effective $\CB{2}$ interaction is purely repulsive. It was shown for composite fermions ($\CF{}$s) that a flattening of the bare $1/r$ potential leads to attractive inter-$\CF{}$ interactions \cite{Scarola2000}. By analogy we speculate that for our extremely flat potential \eqref{eq:effPot} the inter-$\CB{2}$ interactions becomes attractive as well, which will indeed lead to $\CB{2}$-pairing \cite{RICE1988}. We also note that even our bare potential has a small attractive component in momentum-space. Whether or why the 
pairing symmetry 
should be $d$-wave as required for the Hf will be devoted to future work.

We conclude this paragraph by noting that this parameter regime,
which may be easier accessible experimentally than the low-filling regimes discussed above, is potentially a very interesting one. 
In order to make more definite statements further studies are needed however.

%%%%%%%%%%%%%%%%%%%%%%%%%%%%%%%%%%%%%%%%%%%%%%%%%%%%%%%%%%%%%%%%%%%%%%%%%%%
%%%%%%%%%%%%%%%%%%%%%%%%%%%%%%%%%%%%%%%%%%%%%%%%%%%%%%%%%%%%%%%%%%%%%%%%%%%
\section{summary}
%%%%%%%%%%%%%%%%%%%%%%%%%%%%%%%%%%%%%%%%%%%%%%%%%%%%%%%%%%%%%%%%%%%%%%%%%%%
%%%%%%%%%%%%%%%%%%%%%%%%%%%%%%%%%%%%%%%%%%%%%%%%%%%%%%%%%%%%%%%%%%%%%%%%%%%

\begin{figure}[hbt]
\epsfig{file=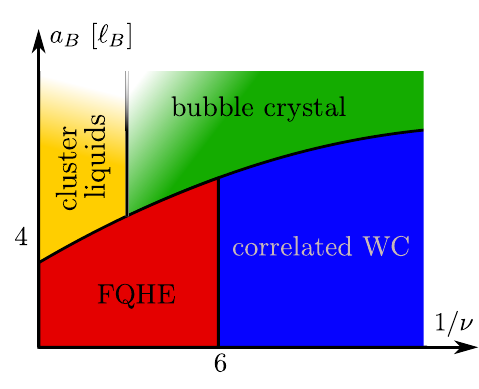, width=0.6\columnwidth}
\caption{(color online). Qualitative form of the LLL phase diagram.}
\label{fig7}
\end{figure}

Summarizing, we have shown that Rydberg-dressed Bose gases can give rise to extremely strong interactions and a variety of interesting correlated phases not found in the standard FQH physics of the LLL. This has two reasons: the rapid fall-off of the interaction potential with distance and the competition of two length scales, $a_B$ and $\ell_B$. A qualitative picture of the LLL phase diagram is shown in fig.\ref{fig7}. In the limit of pure van-der-Waals interactions ($a_B/\ell_B \rightarrow 0$) the $\nu=1/2,1/4$ Laughlin states are exact ground states. Although we mainly discussed filling fractions with even values of $1/\nu$ the entire region is denoted by ``FQHE`` since we expect the standard bosonic FQH physics \cite{Regnault2003} to hold for all fractional fillings larger than $\nu=1/6$. For $\nu \leq 1/6$ we find correlated Wigner crystal ground states, while a noncorrelated Wigner crystal ansatz predicts crystallization only at $\nu=1/12$. For $a_B \gtrsim \sqrt{4/\nu} \ell_B$ and $\nu \leq 1/4$ a 
transition to a bubble crystal is expected. Finally, for larger fillings and blockade radii the nature of the ground state is an open question, and for filling $\nu=1/2$  we speculated on a connection to interesting cluster liquid states in particular the Haffnian \cite{Green2001}.

%%%%%%%%%%%%%%%%%%%%%%%%%%%%%%%%%%%%%%%%%%%%%%%%%%%	%fig 6 was previously here as well;
%%%%%%%%%%%%%%%%%%%%%%%%%%%%%%%%%%%%%%%%%%%%%%%%%%%%%%%%

We thank J. Otterbach for helpful discussions and M. Baranov for drawing our attention to the crystal stability analysis. F.G. acknowledges financial support from the the Graduate School of Excellence MAINZ.

\bibliographystyle{apsrev4-1}

%Merlin.mbs v4.21 2009-07-09.
%

%%%%%%%%%%%%%%%%%%%%%%%%%%%%%%%%%%%%%%%%%%%%%%%%%%%%%%%%%%%%%%%%%%%%%%%%%%%%
%%%%%%%%%%%%%%%%%%%%%%%%%%%%%%%%%%%%%%%%%%%%%%%%%%%%%%%%%%%%%%%%%%%%%%%%%%%%
\section*{Appendix: Exact diagonalization}
%%%%%%%%%%%%%%%%%%%%%%%%%%%%%%%%%%%%%%%%%%%%%%%%%%%%%%%%%%%%%%%%%%%%%%%%%%%%
%%%%%%%%%%%%%%%%%%%%%%%%%%%%%%%%%%%%%%%%%%%%%%%%%%%%%%%%%%%%%%%%%%%%%%%%%%%%

\label{sec:ED}
In addition to variational calculations we performed state-of-the-art exact diagonalization (ED) studies. 
To minimize the role of finite-size effects we did all our systematic investigations in spherical geometry. We work on standard desktop 
computers and have excess up to the following particle numbers:
\begin{center}
\begin{tabular}{c||c|c|c|c|c}
 $\nu$       & $1$ & $1/2$ & $1/4$ & $1/6$ & $1/8$ \\ \hline
 $N_{\max}$ & $14$  & $10$ & $7$    & $6$ &   $6$\\ \hline
dim. & $194668$ &  $246448$ & $48417$ & $32134$ & $118765$
\end{tabular}
\end{center}
The dimension of the full Hilbertspace containing all angular momentum multiplets is given, since we do not 
explicitly exploit that $[\mathcal{H},\vec{L}_{\text{tot}}^2]=0$ in the numerics. 
In the numerics the filling fraction $\nu$ is adjusted by changing the magnetic flux $N_{\Phi}$ 
through the surface of the sphere. For $\nu=1/2,1/4,...$ Laughlin states
\begin{equation*}
 N_\Phi = \frac{N-1}{\nu}.
\end{equation*}
We eliminate leading-order finite size effects by rescaling the magnetic length \cite{DAMBRUMENIL1989},
\begin{equation*}
 \ell_B \rightarrow \ell_B^{\infty} = \sqrt{\frac{N_\Phi \nu}{N}} \ell_B.
\end{equation*}
In simulations $a_B$ must be chosen small enough since when the blockade area fills half a sphere, 
i.e. when $a_B\geq \pi \sqrt{N_{\Phi}/8} \ell_B$, we expect the formation of two clusters on opposite poles. We indeed 
found large overlaps to two-cluster trial wavefunctions in this case and no conclusions about the thermodynamic limit can be drawn.

In disc geometry we only performed systematic studies for small $a_B \lesssim \ell_B$ and found similar 
results as in spherical geometry. The accessible system sizes are the same as above, although requiring slightly more CPU time.

\end{document}